\documentclass[aps,pre,onecolumn,superscriptaddress]{revtex4}

\usepackage{t1enc}
\usepackage{amsmath}
\usepackage{graphicx}
\usepackage{epsfig}
\usepackage{psfrag}
\usepackage{times}

\newcommand{\beq}{\begin{equation}}
\newcommand{\eeq}{\end{equation}}
\newcommand{\beqa}{\begin{eqnarray}}
\newcommand{\eeqa}{\end{eqnarray}}
\newcommand{\vecx}{\vec{x}}
\newcommand{\veck}{\vec{k}}
\newcommand{\ddk}{\frac{d^d k}{(2\pi)^d}}
\newcommand{\tf}{\tilde{f}}
\newcommand{\tk}{\tilde{K}}
\newcommand{\hatphi}{\hat{\phi}}

\begin{document}

\title{Langevin Dynamics of Fluctuation Induced First Order Phase Transitions: 
self consistent Hartree Approximation}

\author{Roberto Mulet}
\affiliation{``Henri-Poincar\'e-Group'' of Complex Systems and
Department of Theoretical Physics, Physics Faculty, University of
Havana, La Habana, CP 10400, Cuba}
\altaffiliation{Research Associate of the
Abdus Salam International Centre for Theoretical Physics}
\author{Daniel A. Stariolo}
\affiliation{Departamento de Física,
             Universidade Federal do Rio Grande do Sul,
             CP 15051, 91501-970, Porto Alegre, Brazil}
\altaffiliation{Research Associate of the
Abdus Salam International Centre for Theoretical Physics}
\date{\today}

\begin{abstract}
 The Langevin dynamics of a system exhibiting a Fluctuation Induced First Order Phase Transition is
 solved within the self consistent Hartree Approximation. Competition between interactions at short
and long length scales gives rise to spatial modulations in the order parameter, like stripes in 2d
and lamellae in 3d. We show that when the time scale of observation is small compared with
 the time needed to the formation of modulated structures, the dynamics is
 dominated by a standard ferromagnetic contribution plus a correction
 term. However, once these structures are formed, the long time dynamics is
 no longer pure ferromagnetic. After a quench from a disordered state to low temperatures the
system develops growing domains of stripes (lamellae). Due to the character of the transition, the paramagnetic
phase is metastable at all finite temperatures, and the correlation length diverges only at $T=0$.
Consequently, the temperature is a relevant
 variable, for $T>0$ the system exhibits interrupted aging while for
 $T=0$ the system ages for all time scales. Furthermore,
 for all $T$, the exponent associated with the
 aging phenomena is independent of the dimension of the system.
\end{abstract}

\pacs{}

\maketitle

\section{Introduction}
Type-II superconductors~\cite{TrStAxNaUc1995}, doped Mott insulators~\cite{KiFrEm1998}, ultrathin magnetic
films\cite{PoVaPe2003,DeMaWh2000,Ja2004}, lipid monolayers~\cite{SeAn1995}, Raleigh-Benard convection
~\cite{HoSw1995},
quantum Hall systems~\cite{BaFrKiOg2002}, are all systems that under appropriate conditions present stable phases 
characterized by the presence of modulated structures. The existence of these modulated structures is well
understood on the basis of the Fluctuation Induced First Order Phase Transition Theory (FIFOT), first
 developed by Brazovskii
~\cite{Br1975}. This scenario predicts that systems in which the spectrum of fluctuations has a minimum in
a shell in reciprocal space at a non 
zero wave vector, undergo a first order phase transition driven by fluctuations, in contrast to the second order 
transition predicted by mean field theory. 
Moreover, the strong degeneracy in the space of fluctuations induces the existence of
many metastable structures at low temperatures, and since the experimentally observed structures are in general 
metastable, dynamical effects become very important. Unfortunately, the dynamical behavior of these systems is 
far from being understood. 

The existence and stability of metastable structures and the nature of the nucleation processes in the context of the
Brazovskii scenario where first studied by Hohenberg and Swift~\cite{HoSw1995}. They obtained the free energy
barriers to nucleation and the shape and size of critical droplets in the weak coupling limit.
Gross et al.~\cite{GrIgCh2000} compared the predictions of the self
consistent Hartree approximation with direct simulations of the Langevin dynamics, confirming the validity of the
approximation.
 
A classic example where the Brazovskii scenario has got strong support, both theoretically and experimentally,
is in diblock copolymers~\cite{FrBi1989,BaRoFr1990,AnHaTrReCh2005}. These systems have interesting technological
applications as self-assembling patterning media.
Another well known example of this kind of system is the three dimensional Coulomb frustrated ferromagnet. 
Wolynes et al.~\cite{WeScWo2001} have shown, using a replica dynamical mean field theory, that below a characteristic 
temperature, an exponential number of metastable states appears in the system preventing long-range order. 
Furthermore, through Monte Carlo simulations~\cite{GrTaVi2001} and also using a mode coupling analysis for the equilibrium 
Langevin dynamics of the Coulomb frustrated ferromagnet,  Grousson et al.~\cite{GrKrTaVi2002} have found an 
ergodicity breaking 
scenario in agreement with the predictions of Wolynes et al. These results resemble the behavior of many glass former
systems, and two theoretical scenarios have been advocated in order to account for its phenomenology
~\cite{StWo2005,TaKiNuVi2005}. The relevance of the mode
coupling predictions to the dynamics of the system have nevertheless been questioned by  Geissler et al.
\cite{GeRe2004}.

The experimental and theoretical study of thin film magnetic materials have led to similar questions~\cite{GaDo1982,
CaStTa2004,PoVaPe2006}. 
Thin films and quasi-two dimensional magnetic materials have many important technological applications,  
for example in data storage and magnetic sensors~\cite{HuSc1998}. In metallic uniaxial ferromagnetic films grown on a
metallic substrate, like CoCu or FeCu, the system develops spontaneous stripe domains upon cooling below the Curie point.
This is due to the competition between the exchange ferromagnetic short range interaction and the antiferromagnetic dipolar 
one, which is long range~\cite{DeMaWh2000}. In the strong anisotropy or Ising limit, a self consistent Hartree approximation
predicts the presence of a FIFOT for any value of the ratio between the ferro and antiferromagnetic coupling constants
~\cite{CaStTa2004}. Nevertheless, the results from Monte Carlo simulations are far from conclusive, suggesting the presence
of first order transitions only for a restricted range of the ratio of the coupling constants
~\cite{RaReTa2006,CaMiStTa2006}. 
The dynamics of these systems is also far from clear. Early work from Roland and Desai~\cite{RoDe1990}, 
who did simulations of
the Langevin dynamics, concentrated on the early time regime, where modulation in the magnetization sets in. Later work 
with Monte Carlo simulations concentrated in some aspects of the out of equilibrium aging dynamics~\cite{ToTaCa1998,StCa1999} 
and the growing of stripe domains after a quench~\cite{GlTaCaMo2003}. These works reveal a very rich phenomenology, with
the appearance of complex phases reminiscent of liquid crystals, and strong metastability of the dynamics. 
All these facts point to
the necessity of a more systematic study of the dynamical aspects of FIFOT.

In this work we solve the Langevin dynamics of a generic model undergoing a first order phase transition 
driven by fluctuations. To characterize the long time dynamics of the system, we study  
the fluctuation spectrum close to the wave vector $k_0$ representative of the modulated phases.
We solve the dynamical equations in the Hartree approximation and
show that, already within this approximation, the dynamics of the system is very rich
and departs from the usual ferromagnetic case. A key observation is that, as we show in the next section,
the spinodal of the high temperature disordered phase shifts to zero temperature in this approximation,
and this has strong influence on the dynamics after a quench. In agreement with the equilibrium results,
we show that the instability of the disordered phase appears only at $T=0$, where the dynamics changes
qualitatively. Nevertheless, the relaxation at finite temperature is far from trivial, showing the emergence
of domains of stripes, which form a kind of mosaic state on top of the striped equilibrium phase.

The rest of the paper is organized as follows. In section \ref{sec:Model}  we
present the model and show that, in the static self consistent approximation, it undergoes a FIFOT. 
In section \ref{dynamics} we introduce the Langevin dynamics.
In section \ref{sec:Hartree} we present the general procedure to
calculate the dynamical properties of the system in the Hartree approximation. 
Sections \ref{sec:Results} and \ref{sec:Results2} are the core of the paper, where
the results on correlations and responses are presented and discussed. 
Some conclusions are presented in section  \ref{sec:Conc}. 
In two appendices we explain some technical details of the calculations.

\section{A Model with a Fluctuation Induced First Order Transition}
\label{sec:Model}

A classical model that undergoes a FIFOT may be defined by an attractive (ferromagnetic) short range 
interaction plus a competing, long-range repulsive (antiferromagnetic) interaction. In the simplest case of
a scalar field, one can define an effective Landau-Ginzburg Hamiltonian of the form:

\begin{equation}
 {\cal H}[\phi]  = \int d^d x \left[ \frac{1}{2} (\nabla \phi(\vec{x}))^2
+\frac{r}{2}\phi^2(\vec{x})+\frac{u}{4} \phi^4(\vec{x}) \right] + \frac{1}{2\delta} \int d^d x\,d^d x'\
\phi(\vec{x})\,J(\vec{x},\vec{x'})\,\phi(\vecx')
\label{eq:hamiltonian}
\end{equation} 

\noindent where $r<0$ and $u>0$. $J(\vec{x},\vec{x'})=J(|\vec{x}-\vec{x'}|)$ represents a repulsive, isotropic,
long range interaction and $\delta$ 
measures the relative intensity between the attractive and repulsive parts of the Hamiltonian. In the limit $\delta 
\to \infty$ one recovers the ferromagnetic $O(N)$ model (for $N=1$)~\cite{CoLiZa2002,NeBr1990,ChCuYo2005}.

The $(u/4)\,\phi^4$ term introduces a non-linearity that makes an exact solution of the model an impossible task. 
To deal with this non-linearity, one must consider the introduction of some kind of 
perturbative analysis. The simplest resummation scheme is
the self-consistent Hartree approximation, or large $N$ limit. 
It consists of replacing one factor of $\phi^2$ in the $\phi^4$ term
in the Hamiltonian by its average $\left<\phi^2\right>$, to be determined self-consistently. There are six ways of 
choosing the two factors of $\phi$ to be paired in $\left<\phi^2\right>$, so the Hamiltonian in the 
Hartree approximation takes the gaussian form:

\begin{equation}
 {\cal H}[\phi] = \frac{1}{2} \int d^dx \left[  \left( \nabla
\phi(\vec{x}) \right)^2 + r\,\phi^2(\vec{x}) + g\,\phi^2(\vec{x})
\left< \phi^2(\vec{x}) \right> \right] + \frac{1}{2\delta} \int
d^dx\,d^dx'\ \phi(\vec{x})\,J(|\vec{x}-\vec{x'}|)\,\phi(\vec{x'})
\end{equation}

where $g=3u$. Introducing the Fourier transform:
\beqa
\phi(\vecx) & = & \int \frac{d^dk}{(2\pi)^d}\ e^{i \vec{k}\cdot \vec{x}}\ \hat{\phi}(\vec{k})\\
\hat{\phi}(\veck) & = & \int d^dx\ e^{-i \vec{k}\cdot \vec{x}}\ \phi(\vec{x})
\eeqa
the Hamiltonian takes the form
\begin{eqnarray}
 {\cal H}[\phi] &=& \frac{1}{2} \int \frac{d^dk}{(2\pi)^d}\ A(k) \hatphi(\veck) \hatphi(-\veck) + \frac{g}{2}  
\int \frac{d^dk_1}{(2\pi)^d}\ \int \frac{d^dk_2}{(2\pi)^d}\ \int \frac{d^dk_3}{(2\pi)^d}\ 
\hatphi(\veck_1) \hatphi(\veck_2)  C(\veck_3,-\veck_1-\veck_2-\veck_3)\\
&=& \frac{1}{2} \int \frac{d^dk_1}{(2\pi)^d}\ \int \frac{d^dk_2}{(2\pi)^d}\ 
\hatphi(\veck_1) \left[ A(k_1) \delta_{\veck_1,-\veck_2}+
g \int \frac{d^dk}{(2\pi)^d}\   C(\veck,-\veck-\veck_1-\veck_2) \right] \hatphi(\veck_2)\\
& = & \frac{1}{2} \int \frac{d^dk_1}{(2\pi)^d}\ \int \frac{d^dk_2}{(2\pi)^d}\ 
\hatphi(\veck_1)\,A(\veck_1,\veck_2)\,\hatphi(\veck_2)
 \label{Hk2}
\end{eqnarray}
with
\begin{equation}
A(\veck_1,\veck_2) = A(k_1)\; \delta_{\veck_1,-\veck_2} + g\, \int \frac{d^dk}{(2\pi)^d}\   
C(\veck,-\veck-\veck_1-\veck_2)
\end{equation}

In the previous expression $A(k)=r+k^2+\hat{J}(k)/\delta$ and $C(\veck,\veck')=\langle \hatphi(\veck)\hatphi(\veck')
\rangle$. Using that $A(\veck_1,\veck_2)=A(-\veck_1,-\veck_2)$ and that
$C(\veck,\veck') = C_c(\veck,\veck')+m_{\veck}\, m_{\veck'}$, where $m_{\veck}=\langle \hatphi(\veck)\rangle$ and
$C_c(\veck,\veck')$ is the connected correlation function, we get finally the self-consistent Hartree equation for the 
connected correlator:
\begin{equation}
A(\veck_1,\veck_2) = \beta^{-1} C^{-1}_c(\veck_1,\veck_2) = A(k_1)\; \delta_{\veck_1,-\veck_2} +
g\,\int \frac{d^dk}{(2\pi)^d}\  \left( m_{\veck}\, m_{\veck_1+\veck_2-\veck} +
 C_c(\veck,\veck_1+\veck_2-\veck) \right)
 \label{HartreeC}
\end{equation}
where $\beta=1/k_BT$. In the paramagnetic phase, at high temperatures, all the order
parameters $m_{\veck}=0$ and the correlation matrix is diagonal,
i.e. $C_c(\veck,\veck') = S_c(\veck)\; \delta_{\veck,-\veck}$, with $S_c(\veck)$ the static structure factor. 
From Eq.(\ref{HartreeC}) we have
\begin{equation}
 \beta^{-1} S^{-1}_c(\veck) = A(k) + g\, \int \frac{d^dk}{(2\pi)^d}\ S_c(\veck)= r + k^2 +
\frac{J(k)}{\delta} + g\, \int \frac{d^dk}{(2\pi)^d}\  S_c(\veck)
\end{equation}
Introducing the ``renormalized mass'':
\beq
\lambda = r + g\, \int \frac{d^dk}{(2\pi)^d}\  S_c(\veck) \label{rmass},
\eeq
the structure factor becomes
\beq
S_c(\veck) = \frac{T}{\lambda+k^2 +\frac{J(k)}{\delta}}
\label{structurefactor}
\eeq
where we have set $k_B=1$, and the renormalized mass has to be determined self-consistently from:
\begin{equation}
\lambda = r + gT \int \frac{d^dk}{(2\pi)^d}\  \frac{1}{\lambda+k^2 +
\frac{J(k)}{\delta}}
\end{equation}
An instability in this equation may appear when $\lambda=\lambda_c= -(k_0^2+\frac{J(k_0)}{\delta})$, 
where $k_0$ is the wave vector which minimizes $A(k)$. Hence, the spinodal temperature $T^*$ is
determined by the equation
\begin{equation}
\beta^* (r^*-\lambda_c) = -g\,K_d \int_0^\Lambda
\frac{k^{d-1}}{\lambda_c+k^2+ \frac{J(k)}{\delta}}\; dk \label{rcpara}
\end{equation}
where $K_d$ is the surface of a d-dimensional sphere.
The integrand in the right hand side is always positive  and
has a singularity at $k=k_0$. Thus, the instability will be
determined by the leading behavior of that integral, which can be
estimated by expanding the denominator of the integrand around
$k_0$:

\[ \int_0^\Lambda
\frac{k^{d-1}}{\lambda_c+k^2+ \frac{J(k)}{\delta}}\; dk \approx
\int_{k_0-\epsilon}^{k_0+\epsilon} \frac{k^{d-1}}{c(k-k_0)^2} dk =
\int_{-\epsilon}^\epsilon \frac{(k+k_0)^{d-1}}{k^2} dk  = \infty\]

\noindent Therefore, the spinodal temperature always is depressed
to zero, that is, $\beta^* r^* \rightarrow -\infty$. The fact that the isotropic phase is
metastable at any finite temperature, a characteristic of the 
self-consistent nature of the fluctuations included in the model, will have important
consequences on the dynamics after a quench at low temperatures. 
Nevertheless, it can be shown that below a melting temperature,
the true equilibrium phase is a modulated one with characteristic wave vector $k_0$ and that a
 first order phase transition driven by fluctuations takes place~\cite{Br1975}.

\section{Langevin dynamics}
\label{dynamics}

As usual, the Langevin dynamics for the scalar field $\phi(\vecx,t)$ is defined by:

\begin{equation}
\frac{\partial \phi(\vecx,t)}{\partial t}=-
\frac{\delta \cal{H}[\phi]}{\delta \phi(\vecx,t)}+\eta(\vecx,t)
\label{eq:lang-dyn}
\end{equation}

\noindent with, in addition, the following conditions for the thermal noise:

\beqa
\langle \eta(\vec{x},t) \rangle & =  & 0 \\
\langle \eta(\vec{x},t) \eta(\vec{x'},t') \rangle & = & 2 T \delta(\vecx-\vecx') \delta (t-t')
\nonumber
\eeqa

\noindent In this work, we consider uncorrelated initial conditions:

\begin{equation}
< \eta(\vec{x},0) \eta(\vec{x'},0)>=\Delta\ \delta(x-x')
\end{equation}

If $\delta \rightarrow \infty$, the last term in equation (\ref{eq:hamiltonian}) may be neglected 
and we keep only the short range part of the potential. At low temperatures this potential 
has two symmetric minima, $\phi(\vecx)=\pm \sqrt{-\frac{r}{u}}$, where $r < 0$. The 
dynamics of this model is well understood\cite{NeBr1990,CoLiZa2002}, and is determined, below
$T_c$, by the fixed point $T=0$. At variance with the complete model, in the pure ferromagnetic
case there is a continuous phase transition at a critical temperature $T_c$ which, in the Hartree
approximation, is different from the mean field critical temperature, with a lower critical
dimension $d_l=2$. In this case, the dynamics after a sub-critical quench corresponds to an usual domain
growth, with a growth law $L(t) \propto t^{1/2}$, as in standard dynamical models with non-conserved
order parameter.

\section{Self Consistent Hartree Approximation}
\label{sec:Hartree}

\subsection{General Solution}

In our case of interest the dynamical equation (\ref{eq:lang-dyn}) reads:

\begin{equation}
\frac{\partial \phi(\vecx,t)}{\partial t}= \nabla^2 \phi(\vecx,t) -r \phi(\vecx,t)-
u\,\phi^3(x,t) - \frac{1}{2\delta}\int d^d x'\,J(\vecx,\vecx')
\phi(\vecx',t) + \eta(\vecx,t)
\end{equation}

To extend consistently the known results for the equilibrium properties of the system to 
the study of the dynamics we keep the same resummation scheme. As before, in this approximation, the non-linear 
term $\phi^3$ is substituted by  \( 3\langle \phi^2(\vecx,t)\rangle \phi(\vecx,t) \) 
where the average is performed over the initial conditions and noise realizations. 
In such a way we obtain a linear equation in \(\phi\) at the price of introducing a 
new parameter $<\phi^2>$ to be determined self-consistently. To proceed, it is useful to go to Fourier space,
in which we can write:

\beq
\frac{\partial \hat{\phi}(\veck,t)}{\partial t}= -[A(k)+I(t)]\ \hat{\phi}(\veck,t) + \hat{\eta}(\veck,t)
\label{eq:dyn-fou}
\eeq
\noindent 

where

\beqa
\label{eq:noise-k}
\langle \hat{\eta}(\veck,t) \rangle & =  & 0 \\
\langle \hat{\eta}(\veck,t) \hat{\eta}(\veck',t') \rangle & = & 2 T \delta(\veck+\veck') \delta (t-t')
\nonumber
\eeqa

\beq
I(t) = r + g\langle \phi^2(\vecx,t)\rangle 
\label{eq:I-t}
\eeq

\beq
A(\veck)=k^2+ \frac{1}{\delta} \hat{J}(\veck,\veck')
\eeq

with initial conditions:

\beqa
\langle \hat{\phi_0}(\veck) \rangle & =  & 0  \\
\langle \hat{\phi}_0(\veck)\ \hat{\phi}_0(\veck') \rangle & = &(2\pi)^d\, \Delta\,\delta(\veck+\veck') 
\label{eq:ci-k}
\nonumber
\eeqa

From equation (\ref{eq:dyn-fou}) it is easy to see that the 
general solution of the model may be written:
\beq
\hat{\phi}(\veck,t)=\hat{\phi}(\veck,0)\,R(\veck,t,0)+\int_0^t 
R(\veck,t,t')\,\hat{\eta}(\veck,t')\,dt'
\label{eq:phi-t}
\eeq

\noindent where
\beq
R(\veck,t,t')=\frac{Y(t)}{Y(t')}e^{-A(\veck)(t-t')}
\eeq
 and  $Y(t)=e^{\int_0^t dt' I(t')}$. 

Our main task is now to find a solution for $Y(t)$,
a function that encloses the unknown parameter introduced 
in the approximation. Following standard procedures~\cite{CoLiZa2002,NeBr1990} it is easy to show that:
\beq
\frac{d K(t)}{d t} = 2 r K(t)+ 2g \Delta f(t) + 4 g T \int_0^t d t' f(t-t') K(t')
\label{eq:K-t}
\eeq
\noindent where $K(t)=Y^2(t)$ and 
\beq
f(t)=\int \ddk e^{-2 A(\veck) t}. 
\eeq
Equation (\ref{eq:K-t}) may be solve by Laplace
transformation methods. If $\tk(p)$ and $\tf(p)$ are the Laplace
transforms of $K(t)$ and $f(t)$ respectively, then
equation (\ref{eq:K-t}) reduces to:

\beq
\tk(p)=\frac{2 g \Delta \tf(p)+K(0)}{p - 2 r - 4 g T \tf(p)}
\label{eq:laplace}
\eeq

Technically, the problem has been reduced to the calculation of $\tf(p)$, to substitute it in
equation (\ref{eq:laplace}) and to calculate the corresponding Bromwich
integral for $K(t)$. Once $K(t)$ is known, all the dynamical quantities
of the system may be easily calculated from integral relations. 

\section{Calculation of $K(t)$}
\label{sec:Results}

In this, rather technical section, we go through a series of approximations and assumptions,
which allow us to compute the function $K(t)$ in the long time limit.
We start the section presenting the approximation used to manage
$A(\vec{k})$. It keeps the necessary ingredients to model a
fluctuation induced first order phase transition.
We then proceed to the calculation of $f(t)$ and finally $K(t)$, in the long time regime. We
show that the cases $T=0$ and $T>0$ give rise to different physics, in agreement with the
static calculations.

\subsection{Approximation for $A(k)$}

Unfortunately, the analytical calculation of $\tf(p)$ for general $A(\veck)$ is a hopeless task. 
We will simplify it, considering only cases in which  $A(\veck)$ depends on the modulus 
of $\veck$, $A(\veck)=A(k)$ (isotropic interactions). Since we are interested in the long time 
dynamics of the model, and we know that the equilibrium phases
are characterized by the existence of a non-trivial wave vector $k_0 \neq 0$, at which the spectrum
of fluctuations has a maximum, it is then natural to develop $A(k)$ close to $k_0$:
\beq
A(k) = A_0 + \frac{A_2}{2}\,(k-k_0)^2 + {\cal O}[(k-k_0)^3]
\eeq
where 
\beqa
A_0 & = & A(k_0) \nonumber \\
A_2 & = & \left. \frac{d^2A}{dk^2} \right|_{k=k_0} \\
  & etc &
\eeqa

with $A_2>0$. 

Note that, if $t$ is large enough,  this approximation is valid 
not only for models with
long range interactions, but in general is a good starting point to 
study other systems whose spectrum of fluctuations has an isotropic minimum at a non-zero
wave vector. Therefore, the reader must keep in mind that the results
of the next sections are valid in a context more general than the 
one represented by the Hamiltonian (\ref{eq:hamiltonian}). 

From a technical point of view, one may note that $A_0$ is irrelevant 
to the dynamical behavior of the system. In fact,
from  equation (\ref{eq:dyn-fou}), one can easily see that it is equivalent to
a rescaling of $r$, and therefore to a shift in the critical temperature of the
system. Therefore, from now on it will be neglected in our calculations.

\subsection{Results for \( \tf(p)\)  }

With the previous assumptions for $A(k)$, we may write $f(t)$ as:

\beq
f(t) = \frac{2\pi^{d/2}}{\Gamma(d/2)}\int \ddk e^{-A_2 t (k-k_0)^2}
\eeq
\noindent whose Laplace transform becomes
\beq
\tf(p) = \int \ddk \frac{1}{p+A_2(k-k_0)^2}
\eeq

Next we analyze the behavior of this integral when $p \simeq 0$. Adding a cutoff factor in the integrals 
in order to regularize the behavior for large wave vectors:

\beq
\tf(p) = \int \ddk \frac{e^{-(k-k_0)^2/\Lambda^2}}{p+A_2(k-k_0)^2} =
\frac{2\pi^{d/2}}{(2\pi)^d \Gamma(d/2)}\int_0^{\infty} dk\ \frac{k^{d-1}e^{-(k-k_0)^2/\Lambda^2}}
{p+A_2(k-k_0)^2}
\label{eq:f-p}
\eeq

Then, after simple algebra equation (\ref{eq:f-p}) becomes:

\beq
\tf(p) = 
\frac{2\pi^{d/2}}{(2\pi)^d \Gamma(d/2)} \frac{1}{(p A_2)^{1/2}}
\int_{-\left(\frac{A_2}{p}\right)^{1/2}|k_0|}^{\infty}
\frac{dk\,e^{-p k^2/A_2 \Lambda^2}}{1+k^2} 
\left[ k_0+ \left( \frac{p}{A_2}\right)^{1/2} k \right]^{d-1}.
\label{eq:last_fp}
\eeq

\noindent Expanding the binomial inside the integral we obtain the 
following expression for general dimensions:

\beq
\tf(p)  =  \frac{2\pi^{d/2}}{(2\pi)^d \Gamma(d/2)} \frac{1}{A_2} \bigg(\frac{A_2}{p}\bigg)^{1/2}
\sum_{j=0}^{d-1} \left( \begin{array}{c} d-1 \\ j \end{array} \right)
\bigg(\frac{p}{A_2}\bigg)^{j/2} k_0^{d-1-j}
\int_{-\left(\frac{A_2}{p}\right)^{1/2}|k_0|}^{\infty} \frac{dk\,e^{-p k^2/A_2 \Lambda^2}}
{1+k^2}  k^j 
\label{f_p}
\eeq
with $d \in \mathcal{Z}$.

At this point two limit cases are possible and will be treated separately below. If
$(\frac{A_2}{p})^{1/2}|k_0| \rightarrow 0$ the time scale  of
observation is such that the stripes are not completely
formed, $\frac{1}{p}<\frac{1}{k_0^2} \frac{1}{A_2}$. In this limit we
recover the dynamic properties  of the pure ferromagnet for $k_0 \sim 0$.
On the other hand, if $(\frac{A_2}{p})^{1/2}|k_0|\rightarrow \infty$ the stripes are already
formed and the interaction among them will be responsible of the
dynamical properties of the system. Once this, more interesting limit, is taken, it is impossible
to recover the pure ferromagnetic behavior.

\subsubsection{Stripes in formation}

Defining:

\beq
F_j(k_0,p)=\bigg(\frac{p}{A_2}\bigg)^{j/2} k_0^{d-1-j}
\int_{-\left(\frac{A_2}{p}\right)^{1/2}|k_0|}^{\infty} \frac{dk\,e^{-p k^2/A_2 \Lambda^2}}
{1+k^2}  k^j 
\label{eq:def-F}
\eeq
and writing it 
as a Taylor series expansion, for $k_0 \sim 0$: $F_j(k_0,p)=F_j(0,p)+F'_j(0,p) k_0+O(k_0^2)$,
we get:
\beqa
F_j(0,p) = \left\{ \begin{array}{ll}
(\frac{p}{A_2})^{(d-1)/2} \int_0^\infty dk \frac{k^{d-1}}{1+k^2} 
e^{-\frac{p}{A_2} \frac{k^2}{\Lambda^2}} & \textrm{if $j=d-1$} \\ 
& \\
0  & \textrm{otherwise} 
\end{array} \right.
\eeqa
and
\beqa
F'_j(0,p)= \left\{ \begin{array}{ll}
(\frac{p}{A_2})^{(d-2)/2}
\int_0^\infty dk \frac{k^{d-1}}{1+k^2} 
e^{-\frac{p}{A_2} \frac{k^2}{\Lambda^2}} & \textrm{if j=d-2} \\ 
& \\
0 & \textrm{otherwise}
\end{array} \right.
\eeqa

Therefore, up to first order in $k_0$, ${F}(k_0,p)$ becomes:
\beq
F_j(k_0,p) =(\frac{p}{A_2})^{(d-1)/2} 
\int_0^\infty dk \frac{k^{d-1}}{1+k^2} 
e^{-\frac{p}{A_2} \frac{k^2}{\Lambda^2}} + k_0
(\frac{p}{A_2})^{(d-1)/2} 
\int_0^\infty dk \frac{k^{d-1}}{1+k^2} 
e^{-\frac{p}{A_2} \frac{k^2}{\Lambda^2}} ,
\eeq

and $\tf(p)$ may be written as:

\beq
\tf(p)=p^{\frac{d}{2}-1}\biggl[
a  + b (\frac{A_2}{p})^{1/2} k_0 \biggr]
\label{eq:cal-fp1}
\eeq

\noindent with $a=\frac{1}{(4 \pi)^{\frac{d}{2}} A_2}$ and $b=-\frac{1}{(4 \pi)^{d/2} A_2^{3/2}}
\frac{\pi \sec(\frac{\pi d}{2})}{\Gamma(\frac{d}{2})}$. 

From equation (\ref{eq:cal-fp1})  it comes out
that  the presence of modulated phases appear in the dynamics as a correction in $\tf(p)$ to the usual 
ferromagnetic case~\cite{CoLiZa2002}. One must remember, however, that this is true provided the second 
term within the brackets is small, i.e. during the formation of the modulated structures. 

\subsubsection{Stripes formed}

On the other hand, for the limit $(\frac{A_2}{p})^{1/2}|k_0|\rightarrow \infty$, stripes of sizes $1/|k_0|$ 
are already formed, and the dynamical properties of the system are defined by their interactions. The Taylor 
expansion, for $k_0$ finite and $p \rightarrow 0$, is written as: $F_j(k_0,p)=F_j(k_0,0)+F'_j(k_0,0) p+O(p^2)$. 
Then, 

\beqa
F_j(k_0,0)= \left\{
\begin{array}{ll}
k_0^{d-1} \int_{-\infty}^{\infty} \frac{dk}{1+k^2} = \pi k_0^{d-1} & \textrm{if $j=0$}\\ 
& \\
0 & \textrm{otherwise}
\end{array} \right.
\eeqa

\noindent and developing as before the first order derivative with respect
to $p$, and taking the limit $p \rightarrow 0$, $\tf(p)$ becomes: 

\beq
\tf(p)=\frac{ 2 \pi^{d/2}}{ (2 \pi)^d \Gamma(d/2)} 
\frac{1}{A_2} (\frac{A_2}{p})^{1/2} \Bigg( \pi k_0^{d-1}+ A_2 k_0^{d-3} \int_{-\infty}^{\infty} dk \frac{k^2}{1+k^2} 
e^{-\frac{p}{A_2} \frac{k^2}{\Lambda^2}}  -\frac{k_0^{d-2}}{2} 
\sqrt{\frac{p}{A_2}} \Bigg) 
\eeq

This last expression shows that for small $p$, $\tf(p)=a+bp^{-1/2}$ with $a<0$,
independently of the system dimension. This limit was also explicitly calculated for $d=1,2$ and
$3$, confirming the series analysis. The calculations are shown in Appendix~\ref{Appendix-f-p}.

Summarizing this subsection, $\tf(p)$ was calculated in two limiting
cases. In the first case, the system is still evolving and the stripes are not formed. The dynamical properties 
resemble the ones of the pure ferromagnet plus a correction term. Once the stripes are formed, the dynamic 
 changes qualitatively, and  one gets that $\tf(p)=a+bp^{-1/2}$, independently of the dimensionality. 
From now on, we will use this expression in future calculations, and only when necessary we will give explicit 
values for $a$ and $b$.

\subsection{The function $K(t)$}

By definition:

\beq
K(t)=\frac{1}{2\pi i} \int_{\sigma-i\infty}^{\sigma+i\infty} dp\ e^{pt} \ \tk(p).
\eeq
\noindent with $\tk(p)$ defined by (\ref{eq:laplace}), where the regularization factors
in $\tf(p)$ can be disregarded in the long time limit. Then: 

\beq
\tk(p) = \frac{2g\Delta a+2g\Delta b p^{-1/2}+1}{p-2A_0-2r-4gTa-4gTbp^{-1/2}}
\eeq

\noindent Because $\tf(p)=a+b p^{-1/2}$, $\tk(p)$ has a branch point at $p=0$,
and the denominator varies in the domain $(-\infty,\infty)$. 

Simplifying the notation we can write

\beq
\tk(p)=\frac{A+Bp^{-1/2}}{p-C-Dp^{-1/2}}=
              \frac{Ap^{1/2}+B}{p^{3/2}-Cp^{1/2}-D}.
\label{eq:kp}
\eeq

\noindent with $A=1+2g\Delta a$, $B=2g\Delta b$, $C=2r+4gTa$ and $D=4gTb$.

The denominator of equation (\ref{eq:kp}) 
has three poles. Through a careful analysis it is possible to show that one pole 
is real and positive for all
temperatures. The other two are complex conjugate with negative real
part. We have to solve:

\beqa
K(t) & = & \frac{1}{2\pi i} \int_{c-i\infty}^{c+i\infty} dp\ e^{p t} \ 
\frac{Ap^{1/2}+B}{p^{3/2}-Cp^{1/2}-D} \nonumber \\
     & = & \frac{1}{2\pi i} \int_{c-i\infty}^{c+i\infty} dp\ e^{p t} \ 
\frac{Ap^{1/2}+B}{(p^{1/2}-x)(p^{1/2}-z)(p^{1/2}-z^{\ast})},
\label{kt}
\eeqa

\noindent where $x^2 \in {\mathcal R}$, $z^2 \in {\mathcal C}$ and
$(z^2)^{\ast}$ is the complex conjugate of $z^2$. After a lengthy
computation (see Appendix~\ref{Appendix-k-t}), we get:

\beqa
K(t) & = & \frac{2\,x\,(A\,x+B)}{(x-z)(x-z^{\ast})} \ e^{ x^2 t} +
       \frac{2\,z\,(A\,z+B)}{(z-x)(z-z^{\ast})} \ e^{z^2 t} +
       \frac{2\,z^{\ast}\,(A\,z^{\ast}+B)}{(z^{\ast}-x)(z^{\ast}-z)} \
       e^{(z^2)^{\ast} t} \nonumber \\
& - & \frac{1}{\pi} \int_0^{\infty} dr\ e^{-rt}\
\frac{B\,r^{3/2}+(BC-AD)r^{1/2}}{D^2+(r^{3/2}+C\,r^{1/2})^2}
\label{kdete}
\eeqa

Now, one must distinguish carefully the cases $T>0$ and
$T=0$.  Note, for example, that the real parts of the complex poles are negative even for
$T \to 0$, while the real pole is always positive going to zero at
$T=0$, where the physics changes qualitatively. 

\subsubsection{$T>0$}

In this case  one may neglect the contributions from the
complex conjugate poles, since their real parts have decaying
exponential functions. On the other hand, the last integral in
equation (\ref{kdete}) may be
easily estimated noting that:
$(r^{3/2}+C\,r^{1/2})^2=r^3+2\,C\,r^2+C^2\,r$, and that for long times 
$t \to \infty$ the dominant contributions will come from $r \ll 1$.
We end with:

\beq
\frac{BC-AD}{D^2} \int_0^{\infty} dr\ e^{-rt}\ r^{1/2} = \frac{BC-AD}{D^2} \frac
{\sqrt{\pi}}{2\,t^{3/2}}.
\eeq

Therefore,

\beq
K(t)  \approx  \frac{2\,x\,(A\,x+B)}{(x-z)(x-z^{\ast})} \ e^{ x^2 t} 
 -  \frac{BC-AD}{2\sqrt{\pi} D^2} \frac{1}{t^{3/2}}.
\label{kdeteT}
\eeq

\noindent where the last term goes to zero for $t \rightarrow \infty$.

\subsubsection{$T=0$}

In this case one must note that $x(T)\rightarrow T$, therefore the
contribution from the real pole dissapears. At the same time, the
complex poles converge to a single real pole that gives rise to a
decaying exponential function.

Moreover, the expansion used to calculate the last integral in (\ref{kdete})
is not longer valid. Being $D=0$ one finds that, for large $t$:

\beq
B \int_0^\infty dr \frac{e^{-r t}}{r^{\frac{3}{2}} + C r^{\frac{1}{2}}}
\sim \frac{B}{\sqrt{\pi} C t^{\frac{1}{2}}}
\eeq

and
\beq
K(t)=A e^{C t} -  \frac{B}{C \pi}\frac{\Gamma(1/2)}{\sqrt{t}} 
\eeq

\noindent with $C<0$. The first term comes from the limit as $T\to 0$ of the two complex poles 
and is obviously subdominant in this analysis.

Already with these results at hand one must note two important
differences with the usual ferromagnetic coarsening. The first one is
that here the temperature is a relevant variable. While for $T>0$ the
relaxation will be dominated by exponential (paramagnetic) contributions, for $T=0$
the relaxation will be power-like. The second one is that, excluding
irrelevant prefactors, the long time dynamics, for all temperatures, {\em is
independent of the system dimensionality}.

\section{Response and Correlation Functions}
\label{sec:Results2}

In this section we present the main physical results of the paper, regarding
the behavior of correlation and response functions.

As seen in Section (\ref{sec:Hartree}), the two-times response function in Fourier space
is given by:
\beq
R(k,t,t')=\frac{Y(t')}{Y(t)}e^{-A(k)(t-t')}
\label{resp}
\eeq
\noindent with $Y(t)=\sqrt{K(t)}$. Defining the two-times structure factor:
\beq
\langle \phi(\veck,t)\,\phi(\veck',t') \rangle = (2\pi)^d\, \delta(\veck+\veck')\,C(\veck,t,t')
\eeq
and using (\ref{eq:noise-k}), (\ref{eq:I-t}) and (\ref{eq:ci-k}),
 one can show that:
\beq
C(\veck,t,t') = \Delta\, R(\veck,t,0)\,R(\veck,t',0)+2T\, \int_0^{t'} R(\veck,t,s)\,R(\veck,t',s)\,ds
\label{eq:C-ktt}
\eeq

As the physics at finite temperature is different from that at $T=0$, we will analyze both cases 
separately. Also, note that, for $T \neq 0$, the leading contribution to $K(t)$ is exponential, 
and consequently, to leading order, correlations and responses will be stationary, consistent with a paramagnetic
phase. Nevertheless, the algebraic subdominant contribution precludes the presence of a transient
non-stationary dynamics, with time scales that can be large for low enough temperatures. 
Consequently, we will keep the subleading contribution also and analyze its effect on the dynamics,
showing that it leads to interrupted aging in correlations and responses. 

\subsection{$T > 0$}

From equation (\ref{kdeteT}) we find:

\beq
Y(t)=\sqrt{\frac{2\,x\,(A\,x+B)}{(x-z)(x-z^{\ast})} \ e^{ x^2 t} 
 -  \frac{BC-AD}{2\sqrt{\pi} D^2} \frac{1}{t^{3/2}}}
\eeq

\noindent and defining $C_2=\frac{2\,x\,(A\,x+B)}{(x-z)(x-z^{\ast})}$ and
$C_1= -  \frac{BC-AD}{2\sqrt{\pi} D^2}$ to simplify the notation, one
gets for large $t$:

\beq
Y(t)= C_1^{\frac{1}{2}} e^{\frac{1}{2} x^2 t} (1+ \frac{C_2}{2 C_1}
\frac{e^{-x^2 t}}{t^{\frac{3}{2}}})
\label{eq:yt}
\eeq

Then, the response function for $T>0$ is:

\beq
R(k,t,t')= e^{-\frac{1}{2}[x^2+A_2(k-k_0)^2](t-t')} 
               \left[ 1+ \frac{C_2}{2C_1} \frac{e^{-x^2 t'}}{t'^{3/2}} \right]
\label{eq:resp-tt-T}
\eeq

The two-time correlation function is given by equation
(\ref{eq:C-ktt}). Then, using (\ref{eq:resp-tt-T}) and  
defining $B(k)=x^2+A_2(k-k_0)^2$ we get:

\beq
C(k,t,t')= \Delta\, e^{-\frac{1}{2}B(k)(t+t')} + 2T\, e^{-\frac{1}{2} B(k)(t+t')}
 \int_0^{t'} ds\ 
e^{-\frac{1}{2}B(k)s} \left( 1+\frac{C_2}{2C_1} \frac{e^{-x^2 s}}{s^{3/2}} \right)^2
\label{eq:Cktt}
\eeq

\noindent Performing the integration in (\ref{eq:Cktt}) and setting $\tau=t-t'$ one gets:



\beq
C(k,\tau,t')= e^{-\frac{1}{2}B(k)\tau} \left[ \frac{2 T}{B(k)} +
\left(\Delta-\frac{2 T}{B(k)} \right)
e^{-B(k) t'}- \frac{C_2}{C_1}
\frac{e^{-\frac{1}{2}x^2 t'} }{\sqrt{t'}} \right]
\label{eq:C-kkt-Ttau}
\eeq

The second term within parenthesis goes rapidly to zero, while the
$\sqrt{t'}$ in the third term reflects the presence of interrupted
aging in the system. On the other hand, keeping the stationary part
and for $t=t'$, we obtain the static structure factor:

\beq
C(\veck) = \lim_{t \to \infty} C(\veck,t) = \frac{2T}{B(k)} =
\frac{2T}{x^2+A_2(k-k_0)^2}
\label{eq:corrTgt0}
\eeq

\noindent which, as expected, shows a characteristic peak at $k=k_0 $.

The correlation function in real space is given by:

\beq
C(\vecx) = \int_{-\infty}^{\infty} \ddk \ C(\veck)\,e^{i \veck \cdot \vecx}
\eeq

In $d$ dimensions:
\beq
C(r) = \frac{1}{(2\pi)^{d/2}} \int_0^{\infty} \left(\frac{1}{kr}\right)^{\frac{d-2}{2}}\,
J_{\frac{d-2}{2}}(kr)\,\frac{k^{d-1}\,dk}{x^2+A_2(k-k_0)^2},
\eeq
where $J_{\nu}(x)$ is a Bessel function of the first kind. In the limit $kr \to \infty$,
\beq
C(r) \approx \frac{1}{(2\pi)^{d/2}} \int_0^{\infty} \left(\frac{1}{kr}\right)^{\frac{d-2}{2}}\, 
\sqrt{\frac{2}{\pi kr}} \cos{\left(kr-\frac{(d-1)\pi}{4}\right)}\,\frac{k^{d-1}\,dk}{x^2+A_2(k-k_0)^2}
\eeq
This integral can be solved using the theorem of residues in the complex plane. The final result is:
\beq
C(r) \propto \cos{(k_0r-\psi)}\ \frac{e^{-r/\xi}}{r^{\frac{d-1}{2}}}.
\label{eq:statcorr}
\eeq
where $\psi = \frac{(d-1)\pi}{4}-\tan^{-1}{\left(\frac{(d-1)x}{2k_0\,\sqrt{A_2}}\right)}$. We see
the presence of a correlation length and a modulation length.
The correlation length is given by

\beq
\xi(T) = \sqrt{\frac{A_2}{x^2}}
\eeq

As mentioned before, at low temperatures, $x(T)\propto T$, and consequently 
the correlation length diverges at $T=0$ as $\xi(T) \propto 1/T$. Then, one can conclude that
 after a quench from the disordered phase to a very low
temperature the system breaks into regions inside which there is modulated order, or stripes. 
No long range order is observed. A transition is approached at $T=0$, where the correlation
length diverges and stripe order sets in. In this respect, it is
 interesting to see also what happens with the correlation function at
 $T=0$.

\subsection{$T=0$}

For $T=0$ we may neglect the decaying exponential and simple algebra
gives:

\beq
R(k,t,t')= (\frac{t}{t'})^{\frac{1}{4}} e^{-\frac{1}{2}[x^2+A_2(k-k_0)^2](t-t')} 
\label{eq:resp-tt-0}
\eeq

Substituting (\ref{eq:resp-tt-0}) in (\ref{eq:C-ktt}) it is easy to
prove that

\beq
C(k,t,t')= \Delta R(k,t,0) R(k,t',0)= \Delta ( t t')^{\frac{1}{4}} 
e^{-\frac{A_2}{2}(k-k_0)^2(t+t')} 
\eeq

\noindent and making $t=t'$ we obtain 

\beq
C(k,t)=  \frac{\Delta}{W}  t^{\frac{1}{2}} e^{-A_2(k-k_0)^2(t)} 
\label{eq:Ckt}
\eeq

\noindent with $W=\frac{B \Gamma(\frac{1}{2})}{\pi |C|}$.

For $d=1$ the spatial correlation function has the form:

\beq
C(r,t)= \frac{\Delta \sqrt{\pi}}{2 W} \cos{( k_0 r)}\  e^{-\frac{r^2}{4 t}}
erf(k_0 \sqrt{t}),
\eeq

\noindent where $erf(x)$ is the error function. This result is consistent, for large $t$,
 with the appearance of modulated structures and long-range order. Nevertheless note that, 
from Eq.(\ref{eq:statcorr}), a critical decay of correlations is observed for $d > 1$.

The
stability of the paramagnetic phase until $T=0$ is obtained also in a purely static calculation
of the phase diagram in the self consistent (Hartree) approximation. Nevertheless, the same
calculation shows that the disordered phase is only metastable, the stripe phase has a lower free
energy below a finite critical temperature and is thus the true thermodynamic equilibrium of the
system. Our calculations show that the Langevin
Dynamics within the Hartree
approximation reproduces this scenario. A quench from a disordered
phase to $T>0$ gives rise to a paramagnetic-like dynamics reflecting
the metastability of this phase. Only at $T=0$, the spinodal is reached, and the system shows
a coarsening non-equilibrium dynamics, with diverging time scales in the thermodynamic limit.

\section{Conclusions}

\label{sec:Conc}

In this work we have calculated within the Hartree approximation the
exact long time dynamics of a model system exhibiting a Fluctuation
Induced First Order Phase Transition. We motivate our work starting with a Hamiltonian with short range 
ferromagnetic interaction and long range antiferromagnetic interactions, but our results are 
valid in general for the long time dynamics of any system exhibiting Fluctuation Induced First Order
 Phase Transitions, provided that the spectrum of the fluctuations is isotropic.
We present explicit expressions 
for one and two time correlation and response functions and show that the dynamics converges to known
static results in the Hartree approximation.
Our results show that
the dynamics of the system may be decomposed in two stages, first
the modulated phases form, and during this stage the dynamics follows
the usual coarsening scenario for a ferromagnetic system: it is dominated by the zero temperature 
fixed point and depends on the dimensionality of the system. Once these
 modulated structures are formed, the dynamics changes qualitatively.
It becomes independent of the system dimension and the temperature becomes a relevant variable.
For $T>0$ the system exhibits interrupted aging and a standard paramagnetic relaxation for large times, 
dominated by the presence of metastable states. At low temperatures domanins of stripes are formed.
At $T=0$ the correlation length diverges and stripe order sets in. The system ages for all the time scales 
following a coarsening dynamics that searches the equilibrium state. Moreover, the exponent associated with this 
aging dynamics is independent of the system dimensionality. 

In this work we have explored only the presence of positional order, through the calculation of the correlations
of the field $\phi(\vecx,t)$. It would be interesting to compute also orientational observables, which are known to be
relevant for these kind of systems, and give rise to nematic-like order~\cite{CaMiStTa2006}. Other interesting
questions that can be addressed starting from the present calculations are the possible nucleation of stripe
phases in the paramagnetic state, which eventually should lead to the first order transition predicted within
the static Hartree approximation. Also the possible presence of freezing in the low temperature dynamics could
be addressed withing a refined approximation, like mode-coupling or the self-consistent screening approximation.

\appendix
\section{}{\label{Appendix-f-p}}

Here we show the explicit calculation of $\tf(p)$ in the limit $p \rightarrow 0$ for $d=1$, 
$d=2$ and $d=3$. We begin from Eq.(\ref{f_p}) in each case, and develop the sums and
integrals.

\subsection*{d=1}

\beq
\tf(p)  =  \frac{1}{\pi (p A_2)^{1/2}}\int_{-\left(\frac{A_2}{p}\right)^{1/2}|k_0|}^{\infty} 
\frac{dk}{1+k^2}e^{-p k^2/A_2 \Lambda^2}
\eeq
Now,
\beqa
\int_{-\left(\frac{A_2}{p}\right)^{1/2}|k_0|}^{\infty} \frac{dk}{1+k^2}
e^{-p k^2/A_2 \Lambda^2} & = & \int_{-\infty}^{\infty} \frac{dk}{1+k^2}
- \int^{-\left(\frac{A_2}{p}\right)^{1/2}|k_0|}_{-\infty} \frac{dk}{1+k^2}
e^{-p k^2/A_2 \Lambda^2} \nonumber \\
 & = & \pi - \int_{\left(\frac{A_2}{p}\right)^{1/2}|k_0|}^{\infty} \frac{dk}{1+k^2}
e^{-p k^2/A_2 \Lambda^2}
\eeqa

The last equality is due to the integrand be an even function. For $p \to 0$ the last integral gives:
\beq
\int_{\left(\frac{A_2}{p}\right)^{1/2}|k_0|}^{\infty} \frac{dk}{1+k^2}
e^{-p k^2/A_2 \Lambda^2} = \left( \frac{p}{A_2} \right)^{1/2}
\frac{e^{-|k_0|^2/\Lambda^2}}{|k_0|} + {\cal O}(p^{3/2})
\eeq
Then,
\beqa
\tf(p) & = & \frac{1}{\pi (p A_2)^{1/2}} \left\{ \pi - \left( \frac{p}{A_2} \right)^{1/2}
\frac{e^{-|k_0|^2/\Lambda^2}}{|k_0|} + {\cal O}(p^{3/2}) \right\} \nonumber \\
 & = & \frac{1}{A_2^{1/2}} p^{-1/2} - \frac{e^{-|k_0|^2/\Lambda^2}}{\pi A_2 |k_0|} + {\cal O}
(p)
\eeqa

\subsection*{d=2}

\beq
\tf(p) = \frac{k_0}{2\pi (p A_2)^{1/2}} 
\int_{-\left(\frac{A_2}{p}\right)^{1/2}|k_0|}^{\infty} \frac{dk}{1+k^2}
e^{-p k^2/A_2 \Lambda^2} 
\left[ 1 + \left( \frac{p}{A_2}\right)^{1/2} \frac{k}{k_0} \right]
\eeq

Here we have to solve two integrals. The first one was already solved for the case $d=1$:

\beq
\int_{-\left(\frac{A_2}{p}\right)^{1/2}|k_0|}^{\infty} \frac{dk}{1+k^2}
e^{-p k^2/A_2 \Lambda^2} = \pi - 
\left( \frac{p}{A_2} \right)^{1/2}
\frac{e^{-|k_0|^2/\Lambda^2}}{|k_0|} + {\cal O}(p^{3/2})
\eeq

The second integral is:

\beq
\int_{-\left(\frac{A_2}{p}\right)^{1/2}|k_0|}^{\infty} \frac{dk\ k}{1+k^2}
e^{-p k^2/A_2 \Lambda^2} = \int_{-\infty}^{\infty} - 
\int_{-\infty}^{-\left(\frac{A_2}{p}\right)^{1/2}|k_0|}=
\int_{\left(\frac{A_2}{p}\right)^{1/2}|k_0|}^{\infty} \frac{dk\ k}{1+k^2}
e^{-p k^2/A_2 \Lambda^2}
\eeq

because the integrand is an odd function. The last integral can be approximated as:
\beq
\int_{\left(\frac{A_2}{p}\right)^{1/2}|k_0|}^{\infty} \frac{dk\ k}{1+k^2}
e^{-p k^2/A_2 \Lambda^2} = 
\int_{\left(\frac{A_2}{p}\right)^{1/2}|k_0|}^{\infty} \frac{dk}{k}
e^{-p k^2/A_2 \Lambda^2} \left[ 1-\frac{1}{k^2}+{\cal O}(k^{-4}) \right]
\eeq
Now

\beqa
\int_{\left(\frac{A_2}{p}\right)^{1/2}|k_0|}^{\infty} \frac{dk}{k}
e^{-p k^2/A_2 \Lambda^2} & = & \frac{1}{2} \Gamma(0,\frac{|k_0|^2}{\Lambda^2}) \nonumber \\
  & = & -\frac{\gamma}{2} - \frac{1}{2}\  \log{\left(\frac{|k_0|^2}{\Lambda^2}\right)} + {\cal O}
\left(\frac{|k_0|^2}{\Lambda^2}\right),
\eeqa
where $\gamma$ is the Euler constant. The other integral
\beqa
\int_{\left(\frac{A_2}{p}\right)^{1/2}|k_0|}^{\infty} \frac{dk}{k^3}
e^{-p k^2/A_2 \Lambda^2} & = & \frac{1}{2}
\left\{ \frac{p\,e^{-|k_0|^2/\Lambda^2}}{A_2 |k_0|^2} - 
\frac{p\,\Gamma(0,\frac{|k_0|^2}{\Lambda^2})}{A_2 \Lambda^2} \right\} \nonumber \\
 & = & \frac{p}{2A_2}
\left\{ \frac{e^{-|k_0|^2/\Lambda^2}}{|k_0|^2} + 
\frac{\gamma}{\Lambda^2} + \frac{1}{\Lambda^2} \log{\left(\frac{|k_0|^2}{\Lambda^2}\right)} 
\right\}
\eeqa
Then,
\beq
\int_{-\left(\frac{A_2}{p}\right)^{1/2}|k_0|}^{\infty} \frac{dk\ k}{1+k^2}
e^{-p k^2/A_2 \Lambda^2} = -\frac{\gamma}{2} - \frac{1}{2}
\log{\left(\frac{|k_0|^2}{\Lambda^2}\right)} - \frac{p}{2A_2}
\frac{e^{-|k_0|^2/\Lambda^2}}{|k_0|^2} + {\cal O}(\Lambda^{-2})
\eeq
Finally,
\beqa
\tf(p) & = & \frac{k_0}{2\pi A_2^{1/2}} \left\{ \pi p^{-1/2} - \frac{1}{A_2^{1/2}}
\frac{e^{-|k_0|^2/\Lambda^2}}{|k_0|} + \frac{1}{k_0 A_2^{1/2}} \left[
-\frac{\gamma}{2} - \frac{1}{2} \log{\left(\frac{|k_0|^2}{\Lambda^2}\right)} +{\cal O}(p)
\right] \right\} \nonumber \\
 & = & \frac{k_0}{2A_2^{1/2}} p^{-1/2} - \frac{e^{-|k_0|^2/\Lambda^2}}{2\pi A_2} -
\frac{\gamma}{4\pi A_2} - {\cal O}(\log{\Lambda^{-2}})
\eeqa

\subsection*{d=3}

In this case

\beqa
\tf(p) & = & \frac{2\pi^{3/2}k_0^2}{(2\pi)^3 \Gamma(3/2)(p A_2)^{1/2}}
\int_{-\left(\frac{A_2}{p}\right)^{1/2}|k_0|}^{\infty} \frac{dk}{1+k^2}
e^{-p k^2/A_2 \Lambda^2} 
\left[ 1 + \left( \frac{p}{A_2}\right)^{1/2} \frac{k}{k_0}+\left( \frac{p}{A_2}\right)
\frac{k^2}{k_0^2} \right] \nonumber \\
 & = & \frac{2\pi^{3/2}k_0^2}{(2\pi)^3 \Gamma(3/2)(p A_2)^{1/2}}
\left\{ \pi - \left( \frac{p}{A_2} \right)^{1/2} \frac{e^{-|k_0|^2/\Lambda^2}}{|k_0|}
+ \frac{1}{k_0}\left(\frac{p}{A_2} \right)^{1/2}\left[ 
-\frac{\gamma}{2} - \frac{1}{2}\  \log{\left(\frac{|k_0|^2}{\Lambda^2}\right)} \right] \right.
+ \nonumber \\
 & & \left. \frac{1}{k_0^2} \left(\frac{p}{A_2}\right)
\int_{-\left(\frac{A_2}{p}\right)^{1/2}|k_0|}^{\infty} \frac{dk\ k^2}{1+k^2}
e^{-p k^2/A_2 \Lambda^2} \right\}
\eeqa

Again, the contribution of the dominant term is of order $p^{-1/2}$, proving that, in
agreement with the series expansion in the text, the behavior of  $\tf(p)$ is independent 
of dimensionality.

\section{}{\label{Appendix-k-t}}

\begin{figure}
\begin{center}
\includegraphics[height=10cm,width=10cm,angle=0]{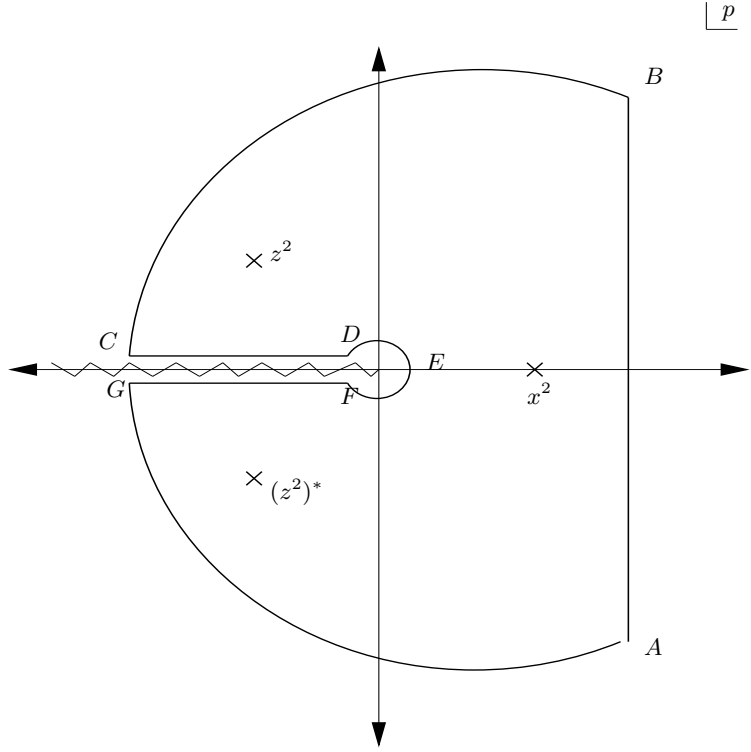}
\caption{Path of integration for the inverse Laplace transform}
\label{complexpath}
\end{center}
\end{figure}

For $\tf(p)==a+b\,p^{-1/2}$ we have

\beqa
K(t) & = & \frac{1}{2\pi i} \int_{c-i\infty}^{c+i\infty} dp\ e^{p t} \ 
\frac{Ap^{1/2}+B}{p^{3/2}-Cp^{1/2}-D} \nonumber \\
     & = & \frac{1}{2\pi i} \int_{c-i\infty}^{c+i\infty} dp\ e^{p t} \ 
\frac{Ap^{1/2}+B}{(p^{1/2}-x)(p^{1/2}-z)(p^{1/2}-z^{\ast})},
\label{kt}
\eeqa
where $x^2 \in {\mathcal R}$, $z^2 \in {\mathcal C}$ and $(z^{\ast})^2$ is the complex conjugate of $z^2$. 
The three poles $x^2$, $z^2$ and $(z^{\ast})^2$ are simple and the residues are:
\beqa
\lim_{p \to x^2} (p-x^2)\ e^{p t}\ \tk(p) & = & 
\frac{(Ax+B)\ e^{x^2 t}\ 2x}{(x-z)(x-z^{\ast})} \nonumber \\
\lim_{p \to z^2} (p-z^2)\ e^{p t}\ \tk(p) & = & 
\frac{(Az+B)\ e^{z^2 t}\ 2z}{(z-x)(z-z^{\ast})} \\
\lim_{p \to (z^{\ast})^2} (p-(z^{\ast})^2)\ e^{p t}\ \tk(p) & = & 
\frac{(Az^{\ast}+B)\ e^{(z^{\ast})^2 t}\ 2z^{\ast}}{(z^{\ast}-x)(z^{\ast}-z)} \nonumber
\eeqa
In order to perform the integral, consider the contour in Fig.(\ref{complexpath}).
\psfrag{A}{$A$}
\psfrag{B}{$B$}
\psfrag{C}{$C$}
\psfrag{D}{$D$}
\psfrag{E}{$E$}
\psfrag{F}{$F$}
\psfrag{G}{$G$}
\psfrag{X}{$x^2$}
\psfrag{Z}{$z^2$}
\psfrag{ZCC}{$(z^2)^{\ast}$}
\psfrag{e}{$p$}
We choose the branch cut to be the negative real axis $(-\infty,0]$. From (\ref{kt}) we conclude that
the paths BC, DEF and GA do not contribute. Then:
\beq
\int_A^B + \int_C^D + \int_F^G = 2\pi i \ \sum Res
\eeq

Define $p=r\,e^{i\theta}$,
\beqa
\int_C^D dp e^{p t}\frac{Ap^{1/2}+B}{p^{3/2}-Cp^{1/2}-D} & = &
\int_{\infty}^0 d(r\,e^{-i\pi})\,e^{r\,e^{i\pi}t} \frac{A\,r^{1/2}\,e^{i\pi/2} + B}
{r^{3/2}\,e^{3i\pi/2}-C\,r^{1/2}\,e^{i\pi/2}-D} \nonumber \\
 & = & -\int_{\infty}^0 dr\ e^{-rt} \frac{B+iA\,r^{1/2}}{-i\,r^{3/2}-i\,C\,r^{1/2}-D} \\
\int_F^G dp e^{p t}\frac{Ap^{1/2}+B}{p^{3/2}-Cp^{1/2}-D} & = &
\int_F^G d(r\,e^{-i\pi})\,e^{r\,e^{-i\pi}t} \frac{A\,r^{1/2}\,e^{-i\pi/2} + B}
{r^{3/2}\,e^{-3i\pi/2}-C\,r^{1/2}\,e^{-i\pi/2}-D} \nonumber \\
 & = & -\int_{0}^{\infty} dr\ e^{-rt} \frac{B-iA\,r^{1/2}}{i\,r^{3/2}+i\,C\,r^{1/2}-D}
\eeqa

Then,
\beqa
\int_C^D + \int_F^G & = & \int_0^\infty dr\ e^{-rt}\ \left\{ \frac{-B+iA\,r^{1/2}}{i(r^{3/2}+C\,r^{1/2})-D}-
\frac{B+iA\,r^{1/2}}{i(r^{3/2}+C\,r^{1/2})+D} \right\} \nonumber \\
& = & \int_0^\infty dr\ e^{-rt}\ \left\{ \frac{B-iA\,r^{1/2}}{D-i(r^{3/2}+C\,r^{1/2})}-
\frac{B+iA\,r^{1/2}}{D+i(r^{3/2}+C\,r^{1/2})} \right\} \nonumber \\
& = & \int_0^\infty dr\ e^{-rt}\ 
\frac{[D+i(r^{3/2}+C\,r^{1/2})](B-iA\,r^{1/2})-[D-i(r^{3/2}+C\,r^{1/2})](B+iA\,r^{1/2})}
{D^2+(r^{3/2}+C\,r^{1/2})^2}
\eeqa
The numerator has the form:
\beqa
xy^{\ast}-x^{\ast}y & = & (\Re x+i\Im x)(\Re y -i\Im y)-(\Re x-i\Im x)(\Re y +i\Im y) \nonumber \\
& = & \Re x \Re y -i\Re x \Im y + i \Re y \Im x + \Im x \Im y -i \Re x \Im y + i \Re y \Im x
-\Re x \Re y - \Im x \Im y \nonumber \\
& = & -2i \Re x \Im y +2i \Re y \Im x
\eeqa
Then,
\beqa
\int_D^C + \int_F^G & = & 2i \int_0^{\infty} dr\ e^{-rt}\ 
\frac{[B(r^{3/2}+C\,r^{1/2})-A\,D\,r^{1/2}]}{D^2+(r^{3/2}+C\,r^{1/2})^2} \nonumber \\
& = & 2i \int_0^{\infty} dr\ e^{-rt}\ 
\frac{B\,r^{3/2}+(BC-AD)r^{1/2}}{D^2+(r^{3/2}+C\,r^{1/2})^2},
\eeqa
and
\beq
K(t) = \sum Res - \frac{1}{2\pi i}\left\{ \int_D^C + \int_F^G \right\}
\eeq

\begin{acknowledgments}
We gratefully aknowledge partial financial support from the Abdus Salam ICTP through grant {\em Net-61,
Latinamerican Network on Slow Dynamics in Complex Systems}. D.A.S. thanks Masayuki Hase for helpful
discussions and CNPq, Brazil, for financial support.
\end{acknowledgments}

\bibliography{ultrathin}

\end{document}